# CDMA Based Interconnect Mechanism for SOPC

Rajesh V[1], Vijaya Kumar P[2]

[1, 2] Department of ECE, Faculty of Engineering & Technology, SRM University
Chennai-603203, India.

[1]*vasireddyrajesh3@gmail.com*
[2]*vijay_at23@rediffmail.com*

**Abstract**
The Network-on-chip (NoC) designs consisting of large pack of Intellectual Property (IP) blocks (cores) on the same silicon die is becoming technically possible nowadays. But, the communication between the IP Cores is the main issue in recent years. This paper presents the design of a Code Division Multiple Access (CDMA) based wrapper interconnect as a component of System on programmable chip (SOPC) builder to communicate between IP cores. In the proposal, only bus lines that carry address and data signals are CDMA coded. CDMA technology has better data integrity, channel continuity, channel isolation, and also mainly it reduces the no.of lines in the bus architecture for transmitting the data from master to slave.
*Keywords: CDMA, On-Chip interconnects, SOPC, Shared Bus Architecture, Encoding and Decoding, Custom component.*

## 1. Introduction

The System on Chip (SoC) consists of a heterogeneous components and functional units that orients towards the specific application domains. The integration of large number of IP cores on a single chip interpolates to the implementation of complex applications. As more and more components are added into an on-chip system, communication issues become more complicated and obstructed. For this issue, Network-on-chip (NoC) [1] is intended to solve the On-Chip communication problems.

Traditional system level design aims at designing reliable single function systems or distributed embedded systems. However, the design complexity exponentially arises as the number of functions on one chip intensifies. Traditional approach is not desirable for such systems. As the number of components on single chip and their performance continue to increase, a shift from computation based to communication based design is becomes mandatory. As a result, the communication architecture plays a major role in the area, performance and energy consumption of overall system.

There are two types of on-chip communication schemes have been considered, namely point-top point (P2P) and bus based communication architectures, P2P communication architectures can provide the utmost in communication performance at the expense of dedicated channels among all the communicating IP pairs. However these architectures suffer from lack of measurability in terms of high complexity, cost and design effort. On the other hand, bus based architectures can connect a few tens of IP cores in a cost-efficient manner by reducing the design complexity and eliminates the dedicated wires required by P2P communication architectures. However, bus-based architectures still fail to satisfy the requirements of future applications mainly due to lack of scalability both in terms of energy and performance.

In contrast to these methods, the Network-on-Chip approach emerged as a promising solution to on chip communication problems. In order to eliminate variance of data transfer latency and complexity incurred by routing issues in a P2P connected NoC [1], an On-Chip network which applies a code division multiple access (CDMA) technique is introduced in this paper. As one of the spread-spectrum techniques, CDMA technique has been widely used in wireless communication systems because it has great bandwidth efficiency and multiple access capability .CDMA technique applies a set of orthogonal codes to encode the data from different users before transmission in shared communication media. Therefore it permits multiple users to use the communication media concurrently by separating data from different users [2]. CDMA NoC is helpful for providing a guaranteed communication service for on-chip system.

SOPC Builder system is used to connect the NIOS processor and CDMA Custom components. Custom components are created by using component editor available in the SOPC Builder [8]. SOPC Builder is a system development tool it enables to define and generate a complete system on programmable chip in less time than traditional and manual integration methods. SOPC is available in ALTERA's QUARTUS II software [7].

Rest of the paper is organized as follows. Section.2 describes about the CDMA Mechanism. Section.3 is about the design solution and related work. Section.4 explains the custom component design in SOPC. Section.5 is about experimental results and Section.6 is the conclusion.

## 2. CDMA Technology

CDMA technology is based on the principle of orthogonality, when a multiple code words are added they do not interfere entirely with each other at every point of time and can be detached without loss of information. Spread code words are comprises of a series of bits which is produced by linear feedback shift register (LFSR). This Code sequences repeats only at every ($2^N$-1) clock cycles, Where N is the number of bits in the shift register. Master encodes the data bits with the specific ($2^N$-1) code bits unique to particular slave. The encoded bits from different master sources can be summed together using parallel counter. These results will ranges from 0 to ($2^N$-1) and are channelized.

CDMA has been widely used for wireless communications. The summed mixture goes through an up-conversion process [3] which translates the frequency to higher band. But for digital bus interconnects, there is no need for up-conversion or down-conversion. However the summation precludes the results to be purely binary.

### 2.1 The CDMA Binary Bus

The Binary CDMA Bus bits are represented as 0 or 1 after the modulation with spreading code and summation. The summations that are less than 0 are ignored [3]. The summer calculate the number chips that are greater than 0 and channelizes the binary equivalent. For 8 processors, only summations from 0 to 8 or 4 equivalent information bits are needed. At the receiver side the original summation from 0 to 8 is reconstructed for the transmitted value P (summed value), the orthogonal summation is (2P-8).

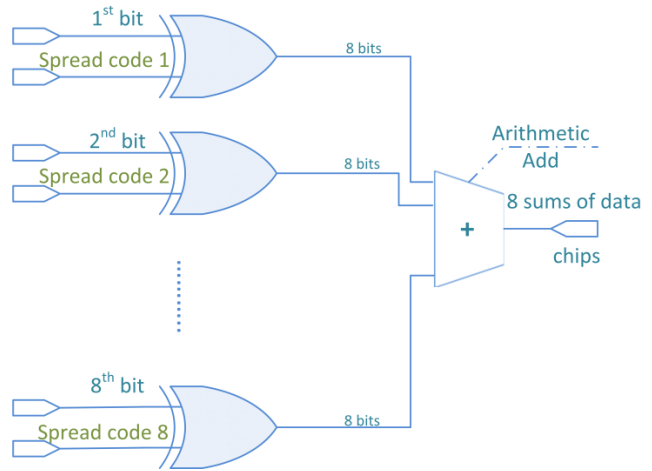

Fig.1 CDMA Encoding Scheme

### 2.2 CDMA Code words

Since the code length is the bandwidth multiplying factor and the number of available codes is the bandwidth dividing factor, it is desirable to have the code word in large number for small number of chips. CDMA transmitter is described in the following Fig.2. Where D represents the data bit and S represents the spread code sequence. To give great orthogonality, taps for the LFSR is taken at 1,2,3,7 registers in an 8-bit shift register and the XOR of these taps are given as the input for the first register. Output of LFSR is taken parallelly at each register, which will look like SIPO shift register.

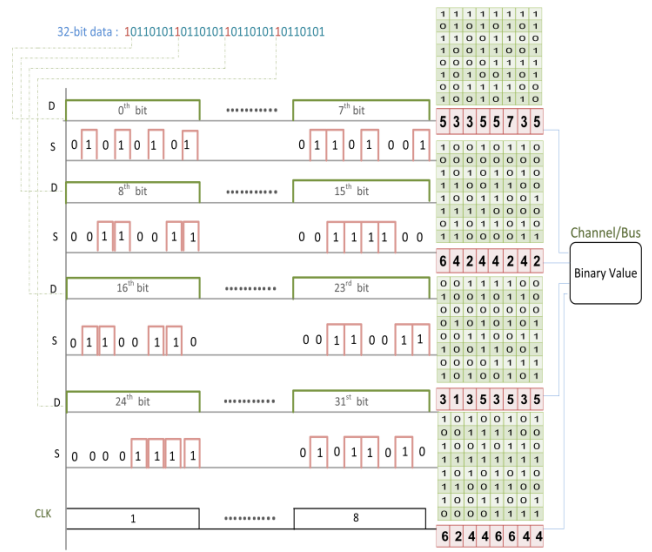

Fig.2.CDMA encoding using Code words of length 8

## 3. Related Work

To send the n-bits of data through the shared bus architecture requires n-bits of lines. But, the CDMA technology reduces the no.of lines required lines for sending n-bits through the shared bus. Let us consider the length of the code word is S and then unencoded buses are of n-bits width then the CDMA coded equivalent buses will be reduced to P=n/s [$\log_2 S + 1$]. For 8-bit spreading code a bus reduction is 50%. Table.1 shows the average bus reduction using the CDMA.

Table.1: No.of Lines Reduced For 'S' and 'N' By CDMA

| ↓S  →N | 8 | 16 | 64 | 128 | 256 |
|---|---|---|---|---|---|
| 4 | 6 | 12 | 48 | 96 | 192 |
| 8 | 4 | 8 | 32 | 64 | 128 |
| 16 | - | 5 | 20 | 40 | 80 |
| 32 | - | - | 12 | 24 | 48 |

CDMA Encoding has been done for 32-bit data using 8-bit length of different spreading code words. Code words are generated by using 8-bit LFSR. The 32-bit data, which comes from the processor, is divided into four batches of 8-bit data. For each clock cycle each single bit of 8-bit data is encoded with 8-bit spreading code. In 32-bit data, $0^{th}$ to $7^{th}$ bits are considered as first batch, $8^{th}$ to $15^{th}$ bits are second batch, 16th to $23^{rd}$ bits are third batch, and similarly $24^{th}$ to $31^{st}$ bits are fourth batch. For every clock cycle, single bit will be taken form each batch and encoded parallelly as shown in the Fig.2. First batch encoding method as an example is shown in Fig.1. After completing 8 clock cycles the encoded data for individual batch is added arithmetically. Maximum Sum of all encoded data will be 8 or 4 equivalent information bits. Further the binary representation of integer value from all batches are serialized and then transmitted through the Channel/Bus. CDMA Encoding scheme uses the principle as shown in the Eq. (1). Where "*SC*" is spreading code, "*D*" is the data bit, and "*C*" is the output. "P" is the summation integer value of encoded data, which is transmitted through Channel/Bus.

$$\sum_{(i,j)=(0,0)}^{(i,j)=(32,7)} SC_i^j \oplus D_i^j = C\ (4bits) \qquad (1)$$

In the case of decoding, for first 8 clock cycles, all the encoded data will be received and decoded as per the formula discussed in Eq.(2). The timing diagrams for encoding and decoding are as shown in the Fig.3 and Fig.4 respectively.

$$D[i] = -2P + N \ if\ codeword[i] = 1$$
$$2P - N \ if\ codeword[i] = 0 \qquad (2)$$

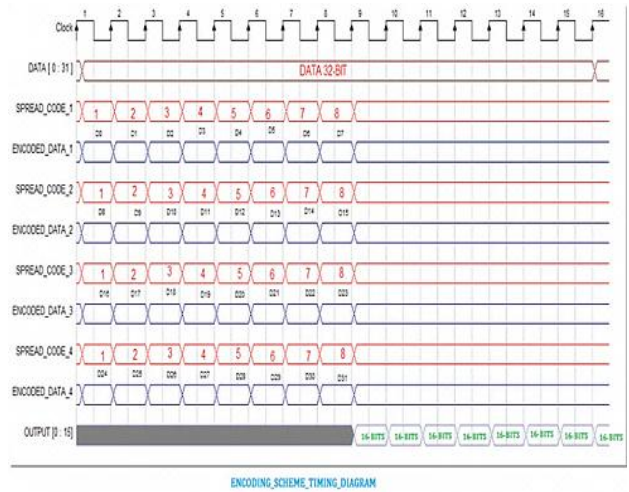

Fig.3 Encoding Scheme timing diagram

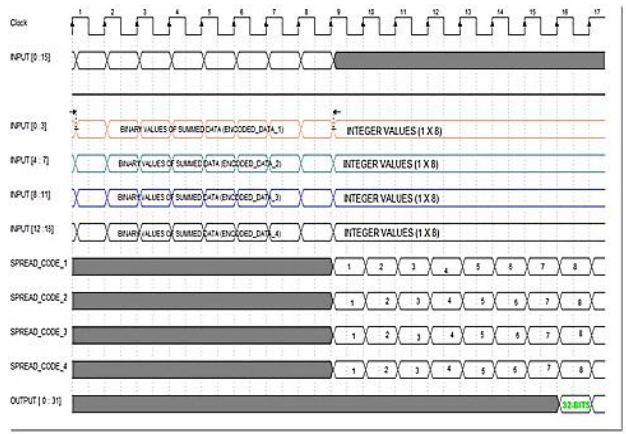

Fig.4 Decoding Scheme timing diagram

## 4. Custom Component Design

The SOPC (System on Programmable Chip) is a powerful tool to create a custom logic design components using component editor. Custom components are connected according to the Avalon Specifications in SOPC Builder. This Avalon defines appropriate interfaces for high speed data, read and write operations of registers and memory. There are seven types of interfaces available in Avalon namely Streaming (ST), Memory Mapped (MM), Conduit, Tristate Conduit, Interrupt, clock and reset interfaces. According to this system design Avalon memory mapped interface is suitable. In order to connect the custom component to the system in must have a slave port. In SOPC, communication is only between Master to slave and vice versa. Two slave ports or two master ports cannot communicate each other.

To develop the custom component, master ports and slave ports should be defined in the form of the following format as shown in Eq.(3).

<Prefix>_<interface name>_<Avalon signal>     (3)

Some of the prefixes and corresponding meanings are shown in the Table.2

Table.2 Prefixes for defining Avalon ports

| Value | Meaning |
|---|---|
| avs | Avalon MM Slave |
| avm | Avalon MM Master |
| ats | Avalon MM Tristate Slave |
| atm | Avalon MM Tristate Master |
| cso | Clock output |
| csi | Clock input |

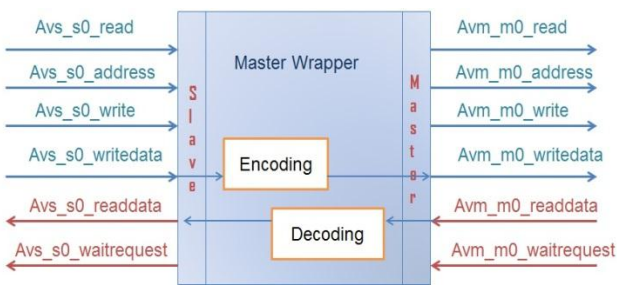

Fig.5 (a) Custom component of CDMA Encoding for SOPC

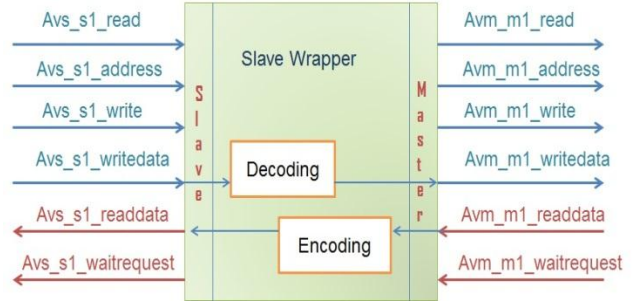

Fig.5 (b) Custom component of CDMA Decoding for SOPC

The signals used in the design of master and slave ports for CDMA components of encoding and decoding are shown in Fig.5 (a) and (b). Signals like "read", "address", "write" are from slave port to master port and given directly. But, "writedata' in the master wrapper is given to Encoding block and in slave wrapper this signal is given to Decoding block. Similarly "readdata" and "waitrequest" are from master to slave port. But, "readdata" in master wrapper is given to Decoding block and in slave wrapper, this signal given to Encoding block.

## 5. Experimental Results

Increasing demand for high-speed on-chip interconnects requires faster links that consumes less power. An on-chip interconnect based on CDMA technique of relatively low complexity, low power and high bandwidth is proposed here and its performance related to design of CDMA are evaluated. The CDMA technique was described in RTL Level using VHDL. For Analysis & Synthesis QUARTUS 10.0v [7] was used and for simulation MODELSIM 10.0c was used. The simulation results for the CDMA Encoding and Decoding is shown in the Fig.6 and Fig.7 respectively.

Resource utilization results of encoding and decoding scheme using CDMA methodology are displayed in Fig.8 and Fig.9. Resources like LEs or ALUTs usage are displayed in graph according to the FPGA device family. X-axis represents Device family.These are CYCLONEIII, CYCLONE IV GX, STRATIX II GX and STRATIX III and Y-axis represents Logic elements or ALUTs values. Fig.9 clearly shows the comparison of logic elements utilized with the logic elements available in the FPGA device family.

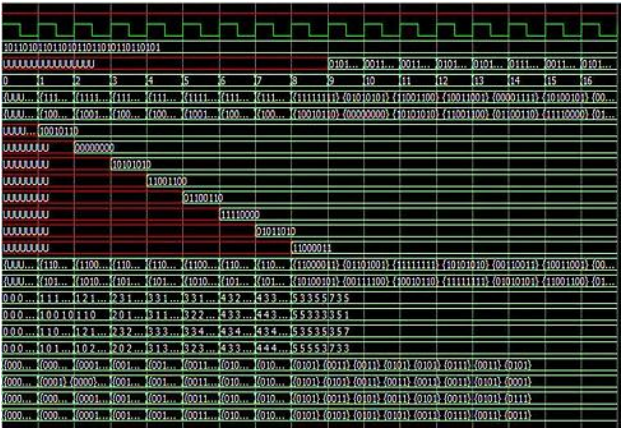

Fig.6 Simulation results of CDMA Encoding Scheme

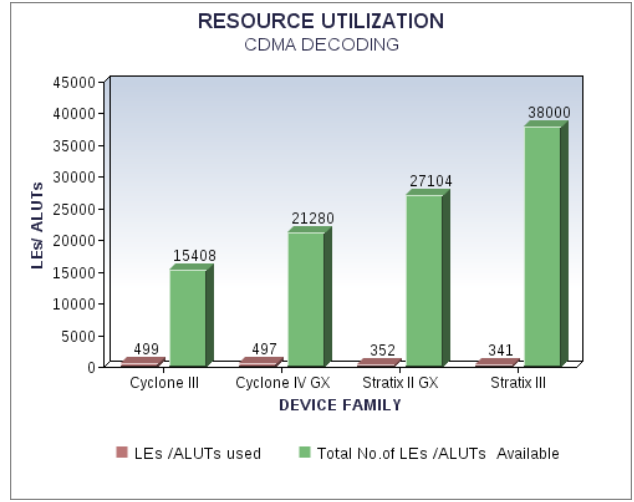

Fig.9 CDMA Decoding Resource utilization

By observing the above results, analysis and synthesis of CDMA encoding using CYCLONE III family uses 677 logic elements and CDMA decoding uses 499LEs out of 15408LEs available which is 4% and 3% respectively. CYCLONE IV GX family uses 638LEs and 497LEs out of 21280 available which is 3% and 2% for encoding and decoding respectively. Similarly for Stratix II GX family uses 410 and 352 combinational ALUTs out of 27104 available. This is 2% and 1% respectively. Stratix III family uses 407 and 341 combinational ALUTs which represents <1% of usage of ALUTs available.

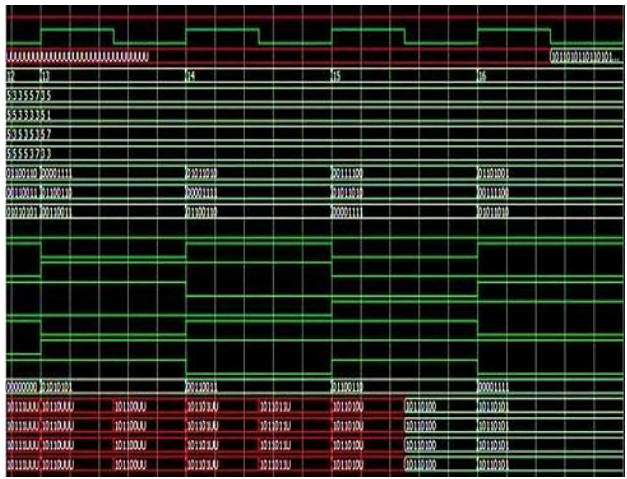

Fig.7 Simulation results of CDMA Decoding Scheme

## 6. Conclusion

In this paper, CDMA encoding and decoding methodologies for the data from the ALTERA's NIOS II [8] processor has been described for SOPC to decrease the no.of lines required to transfer data into the Channel/Bus. This encoding scheme is only applicable to address and the data buses and not for the control lines. Decoding scheme will retrieve the entire data without any loss of information sent from master side. Similarly, designing of custom components or custom IP Core using component editor available in SOPC has been discussed. Simulation results of CDMA ENCODING and DECODING and Resource utilization results of CDMA for different ALTERAs FPGA families are discussed and compared with available resources.

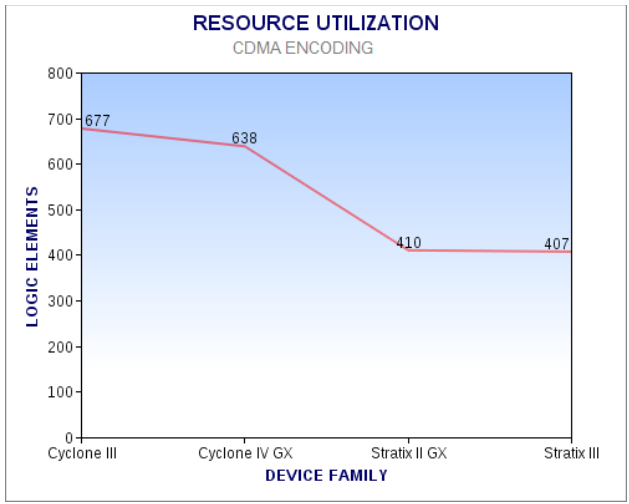

Fig.8 CDMA Encoding Resource utilization

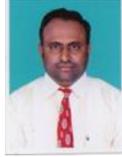

**Mr.P.Vijaya kumar** is working as Asst. Professor (Sr. Grade) in department of ECE, SRM University. He received his M.tech (Applied Electronics) from College of Engineering, Guindy, Anna University, Chennai and B.E (Electronics and communication engg.) from Madras University, Chennai. His research area of interest is Intelligent Signal Processing, Wireless Sensor Networks and Real time computation in SOC. He is life time member of IACSIT, IAENG, ISC, ACEEE and ACM.

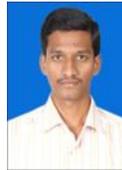

**Mr.V.Rajesh** is pursuing his Master's Degree in Embedded Systems Technology from SRM University, Chennai. He received B.E Degree (Electronics and Communication Engg.) from Anna University, Chennai in 2009.He is the student member of IEEE, ISTE. His current research interests include System on Chip design, embedded systems, Real Time Systems and wireless sensor Network communication.